# Circular Motion of Strings in Cellular Automata, and Other Surprises

Daniel B. Miller and Edward Fredkin

**Abstract**

A two-state, three-dimensional, deterministic, reversible cellular automaton is shown to be capable of approximately circular orbits, wavelike undulations, and particle-like configurations that decay in accordance with a half-life law.

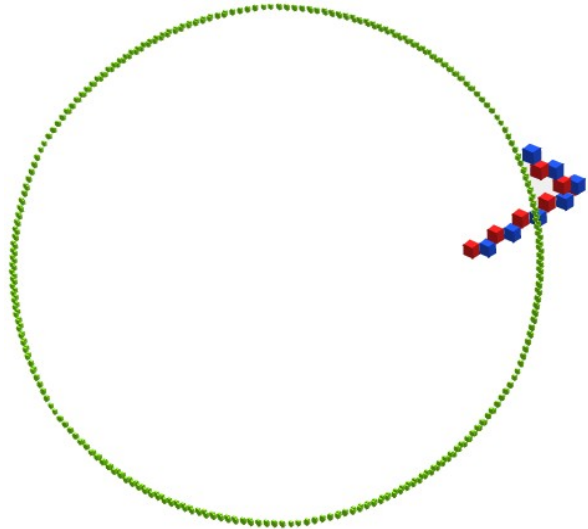

***Fig 1:*** *Busy Boxes CA tracing a quasi-circular path*

## 1 Introduction

In our earlier paper, titled *Two-state, Reversible, Universal Cellular Automata In Three Dimensions* [1], we introduced a novel family of intrinsically reversible CA's based on our (Fredkin's) SALT architecture, where every cell is identified as either even or odd. This can be imagined as a three-dimensional lattice of cubes or 'cells'. Even cells are located at the positions in the lattice where the sum of the three integer coordinates is even; odd cells are at the positions where the coordinate sum is odd. The term SALT comes from the fact that the even and odd positions on the lattice correspond to the positions of sodium and chloride ions in a crystal of table salt.

At any integer time (referred to as a 'step' or 'time step'), each cell in the lattice can either be *up*, which we represent with a 1, or *down,* which we represent with a -1. The rules we allow for a SALT-type Cellular Automaton (CA) always follow this principle: in one time step, the local states of the even cells (whether they are -1 or 1) are used to determine whether or not to the states of pairs of odd cells will be swapped. In the following time step, the states of the odd cells are used to determine whether or not the states of pairs of even cells will be swapped. The pairs of cells whose states we swap are always diagonally situated neighbors in one of three planes perpendicular to the three axes. We continue in this way as long as we wish to evolve the state of the CA. The combination of three orthogonal planes and parity (even or odd) means that SALT models evolve through six distinct steps, as will be explained in detail.

We have researched a number of potential rules within this SALT paradigm. One set of rules in particular was chosen for deeper analysis, and it is this rule, which we refer to as the *Busy Boxes* rule, that produced the results reported in the previously mentioned paper. In this paper, we wish to present some further results based on this SALT variant.

A footnote is included for each configuration, which includes a hyperlink to a website where that configuration's behavior can be visualized in real time. The footnotes also include a list of coordinates of all the 'up' cells for that configuration. The website requires support for the HTML-5 <canvas> element, which is included in recent versions of most popular browsers. Internet Explorer 8 and earlier are not supported.

## 2   Recap of earlier work

In our previous work [*1*], we show that this CA is capable of universal computation, using reversible logic [*2*]. The argument tracks that put forth by Norman Margolus for the Billiard Ball Computer [*3*], showing that a CA that is capable of signal routing and (reversible) logic operations must be computation universal in the Turing sense. We also propose, though we do not provide a proof, that this CA is capable of so-called 'universal construction', in the sense put forward by Von Neumann [*4*].

## 3   New results

Since the publication of our earlier paper introducing this family of Cellular Automata and the Busy Boxes variant, we have made several new discoveries that we think are intriguing and deserving of further study. In this paper, we reintroduce the Busy Boxes cellular automaton, first as a two-dimensional CA, and then in three dimensions. Along the way, we introduce new behaviors not previously encountered, including strings that move along circular paths, strings that show wavelike behavior, and configurations that exhibit long, complex interactions before repeating or 'decaying' into simpler configurations.

## 4   The Rule

To fully explain the rules of this 3D automaton, it is best to first describe a two-dimensional variant. This 2D rule is then applied in all three dimensions in a specific way, which we describe in due course.

The two-dimensional version operates on a flat chessboard-like lattice, with even and odd squares (cells), as defined by the sum of their integer coordinates being even or odd. When operating on one set of cells (even or odd), we make a decision whether to swap the state of two cells situated diagonally. In Fig 2, we show the position of one pair of odd cells to swap, labeled A and B.

The determination as to whether to swap states or not for a given pair of diagonal cells is based on the state of the cells situated in what we refer to as the *Knight's Move* position. These are the cells that can be reached from *both* cells in the swap pair by moving 1 cell over in one axis, and 2 cells in the other axis. Looking again at Fig 2, the Knight's Move positions are labeled C and D. Note that these cells are of opposite parity (or color) to the ones we are considering swapping (in the figure, the cells to swap are odd and colored blue, and the Knight's Move cells are even, colored red.) If the cells to swap are even, their Knight's Move cells are odd, and vice versa.

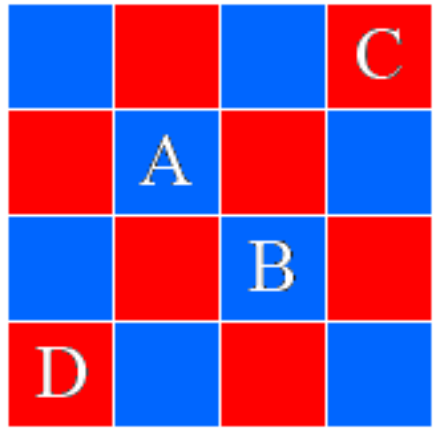

*Fig 2:* *A and B are the cells to be swapped if C and/or D are 'up'. Red cells are even; blue are odd.*

If there is an 'up' cell at one or both of the Knight's Move positions for a given diagonal pair of cells, we will provisionally decide to swap the states of that pair of cells. We apply this rule to every pair of diagonally adjacent odd or even cells in every time step (including both possible diagonal orientations).

## 4.1 Conflicts and the *no swap* rule

One issue we must keep in mind to design a useful CA is that any rule we invent must apply without ambiguities, purely by looking at local information. One problem with our swap rule as stated is that we may end up swapping the states of two cells, and then swap states of another two cells, one of which includes a cell in the first pair. The outcome of such a sequence of steps may depend on the order in which the swaps are done, which is not what we want. We address this issue by applying the rule in the following way, with a caveat to deal with these sorts of conflicts.

First of all, we implement each time step as a two-pass process. In the first pass, we mark every diagonal pair that we intend to swap. In the second pass, we actually perform the swap, but only in the case where there is exactly one unambiguous swap specified for both cells. We refer to this caveat as the *no swap* rule: we will not swap the state of two cells if there is a possible swap specified for either cell in the pair, other than with each other.[1] It is easy to see that this rule avoids the sequencing and locality issue: no matter what order we use to process the cells, the outcome will be the same. This also means that this CA is amenable to being implemented in a completely parallel fashion (albeit requiring a somewhat extended local neighborhood of cells to check).[2]

---

1 Another way of describing the rule is to think of the cells that are 'up' as pieces on a checkerboard. Red pieces are always on the red squares, and black pieces are on the black squares. At every turn, we move either red or black pieces. For each piece of one color, we look for pieces of the other color that are a knight's move away. If we see any such pieces, we consider moving our piece diagonally such that the knight's move piece is still a knight's move from us (there is always exactly one way to do this). We only make this move if A) it is the only move we can make, B) there is no piece in the square we wish to move to, and C) making the move does not put us into a knight's move relationship with any piece other than the one(s) that inspired the move in the first place.

2 To calculate whether a cell will need to be swapped (and in what direction if so) without regard to the disposition of other cells, it is necessary to check a neighborhood of 24 cells distributed within a 7 X 7 square centered around the cell in question. This only needs to be done for 'up' cells, and the calculation can be optimized if branching and early termination are permitted.

# 5   Strings

The 2-dimensional version of this CA has some interesting properties that are worth analyzing before we move into the third dimension. When limited to a single plane, the rules we have chosen allow the formation of series of interconnected 'up' cells, which we refer to as strings. The simplest string is simply a pair of odd and even cells in a Knight's Move relationship, which in our earlier paper we called a 'glider'. These simple 2-cell strings can be seen to 'move' across the lattice in one of four diagonal directions, depending on their initial configuration. Fig 3 shows a string that moves towards the upper right as time progresses.

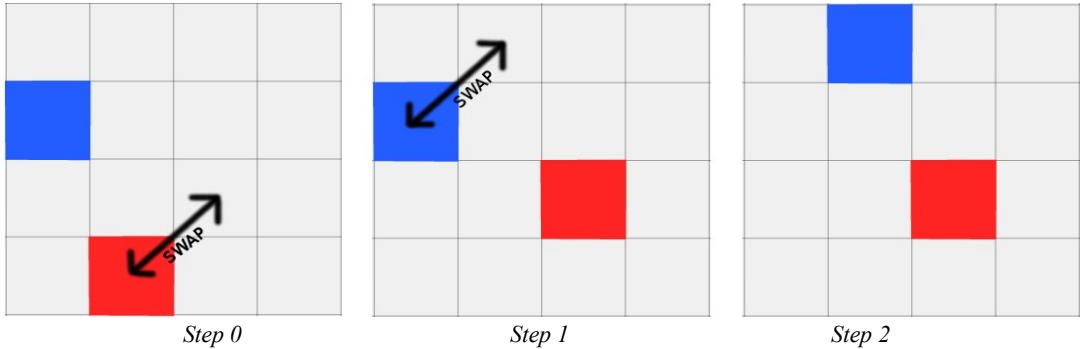

*Step 0*                   *Step 1*                   *Step 2*

**Fig 3**: *'Up' cells are shown as red if even, blue if odd. All other cells are 'down'.*[1]

A 3-cell string can take one of two possible modes -- either stationary or moving[2]. 4-cell strings also have two modes -- fast and slow. Strings can switch between their possible modes, and also reverse direction, depending on how they interact with other cells[3].

As we make strings longer, the number of possible modes increases. We can use these properties to create strings that move at arbitrary speeds. Let's call the fastest possible velocity C, the velocity of a 2-cell string. The 4-cell string at its slowest moves at a velocity of C/3 (see Fig 4)[4]. We can create a C/2 string with 5 cells[5], and a 2/3 C string with 7 cells[6].

It appears that in two dimensions, strings cannot be created or destroyed. The mechanics of our rule set prohibit a string from changing length through the interactions available on a single plane. As we will see, this does not hold in three dimensions.

---

1   http://busyboxes.org/a coordinates: (-1,0,2),(0,0,0)
2   http://busyboxes.org/b coordinates: (-1,0,2),(0,0,0),(2,0,-1),(7,0,-4),(8,0,-6),(9,0,-8)
3   http://busyboxes.org/c coordinates: (-2,0,1),(-3,0,-9),(0,0,0),(2,0,-1),(4,0,-2)
4   http://busyboxes.org/d coordinates: (-1,0,-1),(-3,0,-2),(1,0,0),(3,0,1)
5   http://busyboxes.org/e coordinates: (-1,0,0),(-2,0,-2),(-4,0,-3),(1,0,1),(3,0,2)
6   http://busyboxes.org/f  coordinates: (-2,0,-1),(-3,0,-3),(-5,0,-4),(0,0,0),(1,0,2),(3,0,3),(5,0,4)

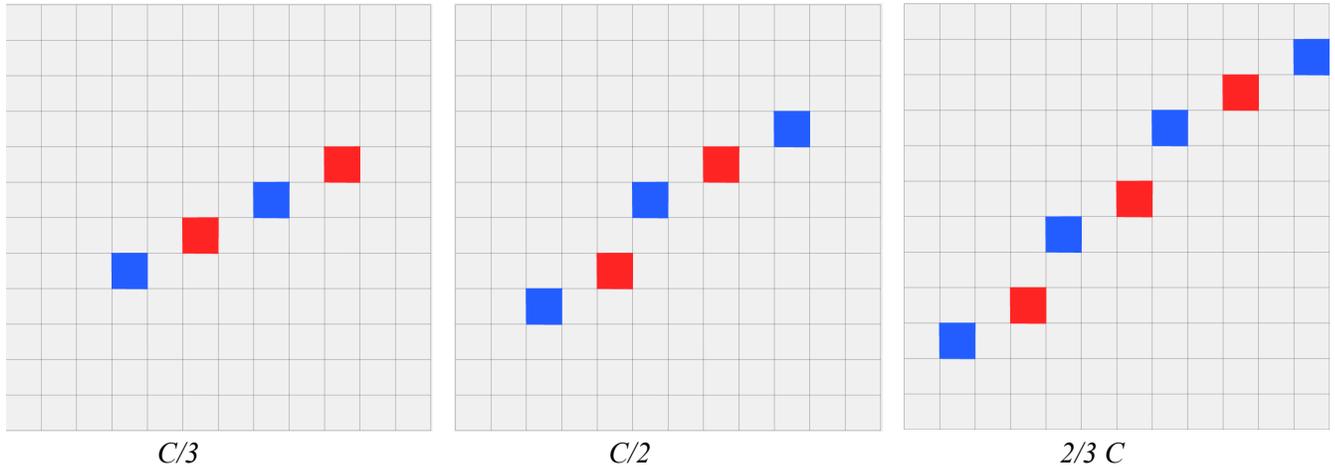

*C/3*           *C/2*           *2/3 C*

**Fig 4:** *strings can move at arbitrary velocities*

## 5.1 Wavelike behavior in strings

We can 'anchor' strings at each end, restricting their ability to move. These constructions show intriguing wavelike behavior, highlighting the potential for reversible CA's to exhibit wavelike state propagation under certain circumstances. In the future, we hope to show similar behavior in three dimensions, either with this CA or with a related CA utilizing somewhat modified rules.

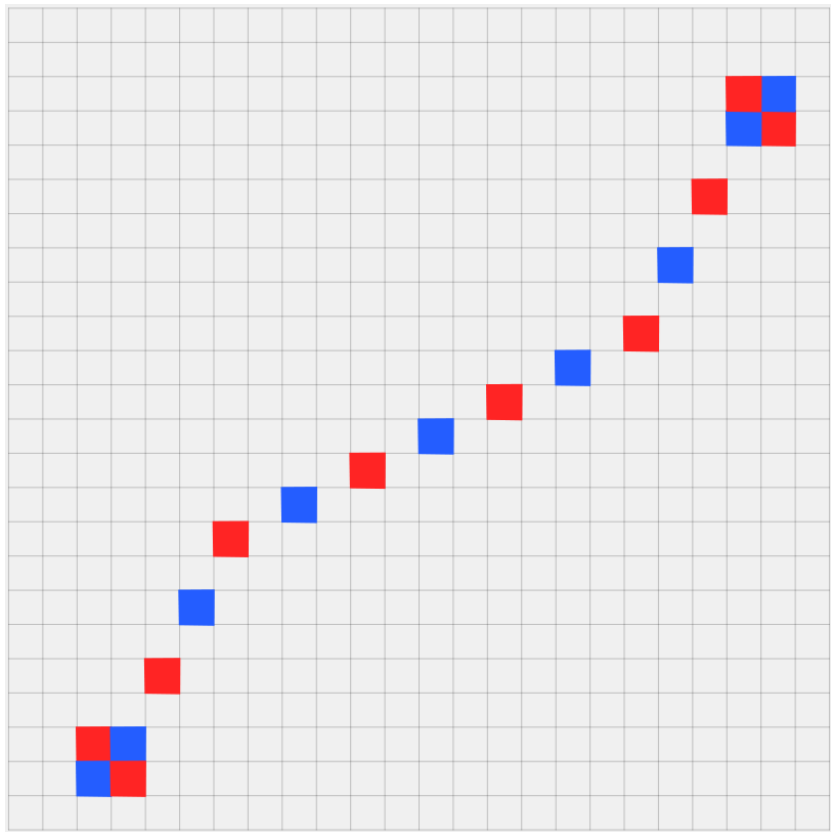

**Fig 5:** *Anchored strings show wavelike behavior*

The basic template for these configurations is a string of cells bound at each end by an 'anchor' configuration of 4 cells[1]. This configuration exhibits an oscillation that can be thought of as the fundamental frequency of this string of 11 cells (not counting the anchor cells), with a period, or cycle time, of 72 steps. A related configuration (Fig 5)[2], has a period of 36 steps, representing the 'first harmonic', or octave, of the previous configuration. Similarly, we can derive a second harmonic at 1/3 of the original period, 24[3]. Finally, we can introduce a disturbance, in the form of a 2-cell string (glider) that repeatedly interacts with the anchored string, causing it to vibrate at multiple frequencies in a chaotic manner[4].

# 6   Three dimensions

Now that we have described the 2D version of the rule, let's apply it in all three dimensions. We do this in a rather obvious way: we apply the 2D rule sequentially in each of the 3 planes that are described by pairs of axes. For each time step, we apply the 2D rule to every layer of cells parallel to one of these 3 planes. Using the traditional X, Y, and Z to define each axis, we actually end up with a 6-phase set of operations, which are as follows:

Phase 0: run 2D rule on XY planes, swap red cells
Phase 1: run 2D rule on YZ planes, swap black cells
Phase 2: run 2D rule on ZX planes, swap red cells
Phase 3: run 2D rule on XY planes, swap black cells
Phase 4: run 2D rule on YZ planes, swap red cells
Phase 5: run 2D rule on ZX planes, swap black cells

There are some edge cases that can be decided in several ways. It's not hard to implement this CA as a virtually infinite lattice, assuming the initial conditions are finitely describable (ie, they can be algorithmically described with a finite-length algorithm). However, in most practical implementations, we choose to bound the volume of the CA to a finite limit, typically a cube of dimension N * N * N, where N is any even positive integer. We can choose to 'wrap' the coordinate space, such that each face of the virtual cube of our lattice is imagined to be adjacent to the opposite, parallel face of the cube. Such a topology is referred to as 'toroidal', as it is topologically equivalent to a 4-dimensional torus. Another option is to prohibit swaps outside the volume, which we call 'mirror' mode. The practical result of mirror mode is that the simple, 2-cell string or glider that is commonly seen in this CA will bounce back from the wall of the lattice volume. Neither of these approaches affect the reversible nature of the CA.

---

1   http://busyboxes.org/g coordinates: (-1,0,3),(-10,0,-10),(-10,0,-11),(-3,0,2),(-4,0,0),(-5,0,-2),(-6,0,-4),(-7,0,-6),(-8,0,-8),(-9,0,-10),(-9,0,-11),
            (1,0,4),(10,0,8),(10,0,9),(3,0,5),(5,0,6),(7,0,7),(9,0,8),(9,0,9)
2   http://busyboxes.org/h coordinates: (-10,0,-10),(-10,0,-11),(-2,0,-2),(-4,0,-3),(-6,0,-4),(-7,0,-6),(-8,0,-8),(-9,0,-10),(-9,0,-11),(0,0,-1),
            (10,0,8),(10,0,9),(2,0,0),(4,0,1),(6,0,2),(7,0,4),(8,0,6),(9,0,8),(9,0,9)
3   http://busyboxes.org/i coordinates: (-1,0,-3),(-10,0,-10),(-10,0,-11),(-3,0,-4),(-5,0,-5),(-7,0,-6),(-8,0,-8),(-9,0,-10),(-9,0,-11),(1,0,-2),
            (10,0,8),(10,0,9),(2,0,0),(3,0,2),(4,0,4),(5,0,6),(7,0,7),(9,0,8),(9,0,9)
4   http://busyboxes.org/j coordinates: (-1,0,3),(-1,0,7),(-10,0,-10),(-10,0,-11),(-3,0,2),(-4,0,0),(-4,0,10),(-5,0,-2),(-6,0,-4),(-7,0,-6),(-8,0,-8),
            (-9,0,-10),(-9,0,-11),(0,0,9),(1,0,4),(10,0,8),(10,0,9),(3,0,5),(5,0,6),(7,0,7),(9,0,8),(9,0,9)

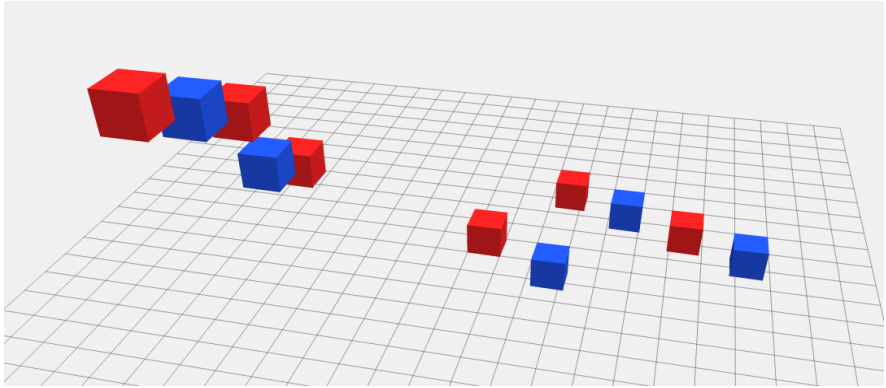

*Fig 6: three strings that move along orthogonal planes[1]*

## 6.1 String motion in 3D

In three dimensions, we can still create the 2D strings shown previously, as long as we restrict their position and orientation so that all the cells lie on a plane perpendicular to one of the three axes. We can therefore create strings that move in any of the 12 directions represented by a combination of -1 or 1 in two axes and zero in the third. Fig 6 shows three strings that move in the directions (1,1,0), (1,0,1), and (0,1,1)[1].

## 6.2 New Directions

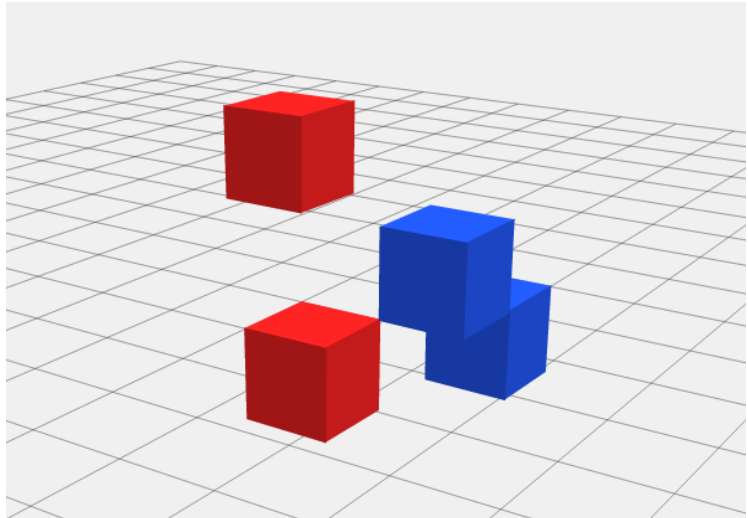

*Fig 7: a glider that moves in one of 8 directions.*

The configuration shown in Fig 7 produces a glider that moves in the (1, 1, -1) direction[2]. All previously known gliders (including the strings mentioned earlier) move in diagonal directions along planes perpendicular to the axes, such as (1, 1, 0), (1, 0, -1) and so on. Permutations of this new glider can produce motion in any of the 8 directions represented by a 1 or -1 for each of the three axes.

---

1  http://busyboxes.org/k coordinates: (-5,3,-2),(-5,4,-4),(-5,6,-5),(-5,7,-7),(-5,8,-9),(0,0,0),(2,-1,0),(2,-1,5),(4,-1,4),(6,-1,3),(8,-1,2)
2  http://busyboxes.org/m coordinates: (-1,-1,0),(-2,1,1),(0,-1,2),(0,0,1)

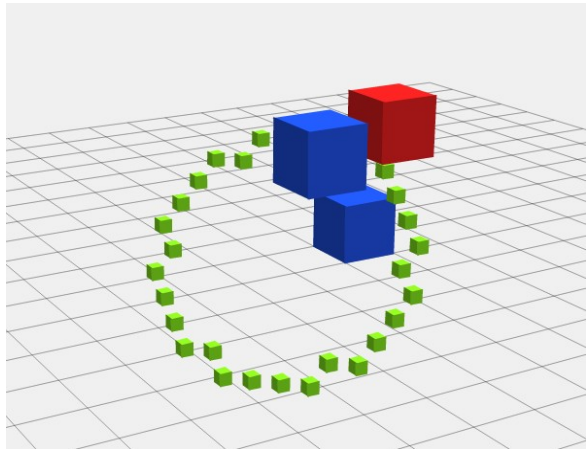
*Fig 8: simplest orbiting string*

# 7 Quasi-circular motion

Returning our attention to strings, there is a peculiar sort of string with a twist that we can create in 3D that exhibits very surprising behavior. Most lattice-based Cellular Automata do not naturally produce circular or orbital patterns, which are presumably a prerequisite to modeling our own universe, the behavior of which appears to be invariant under arbitrary rotations in all axes. However, these 'twisted strings' closely approximate a circular path when plotting the average of their location over time. Circular paths of increasing radius can be described by extending the length of the initial configuration along one plane in an alternating Knight's Move pattern, similar to the 2D strings discussed previously.

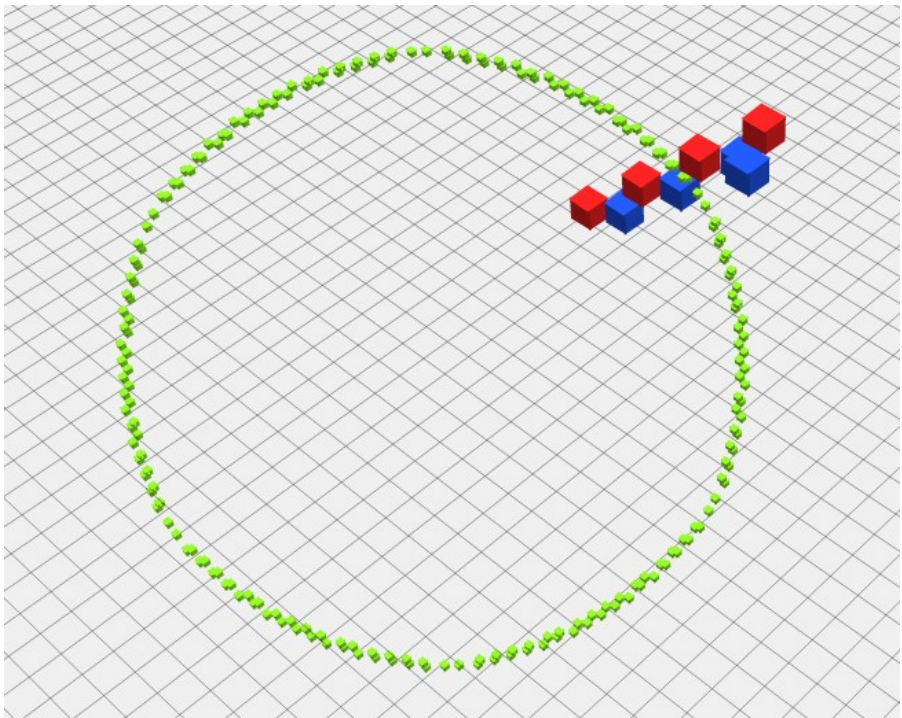
*Fig 9: 8-cell string traces a quasi-circular path (radius 11.32, cycle 330).*

The simplest orbiting string (Fig 8) is created using three cells[1], in a Knight's Move relationship with each other, but in two different axis-aligned planes. The path it traces is roughly hexagonal, and all the points in its path lie on a single plane. As we increase the length of the string (by adding cells diagonally along one of the planes), we also increase the radius of the path it traces. As the radius increases, the string's path (calculated as the average position of each 'up' cell in the string) begins to closely approximate a circle (Fig 9)[2]. We have analyzed the degree to which these paths approximate circles: as the radius increases, the standard deviation from a true circle, plotted as a percentage of the radius, approaches zero (Fig 10). The radius is computed as the average distance of each iteration from the center, which in turn is calculated by taking the average position over the entire path traced by the string. Once the string has completed a full circuit, it is in the same configuration and phase as the initial conditions, and unless interfered with, it will continue to trace this path ad infinitum. Fig 1 shows a 14-cell string that traces an almost circular path in 654 steps.

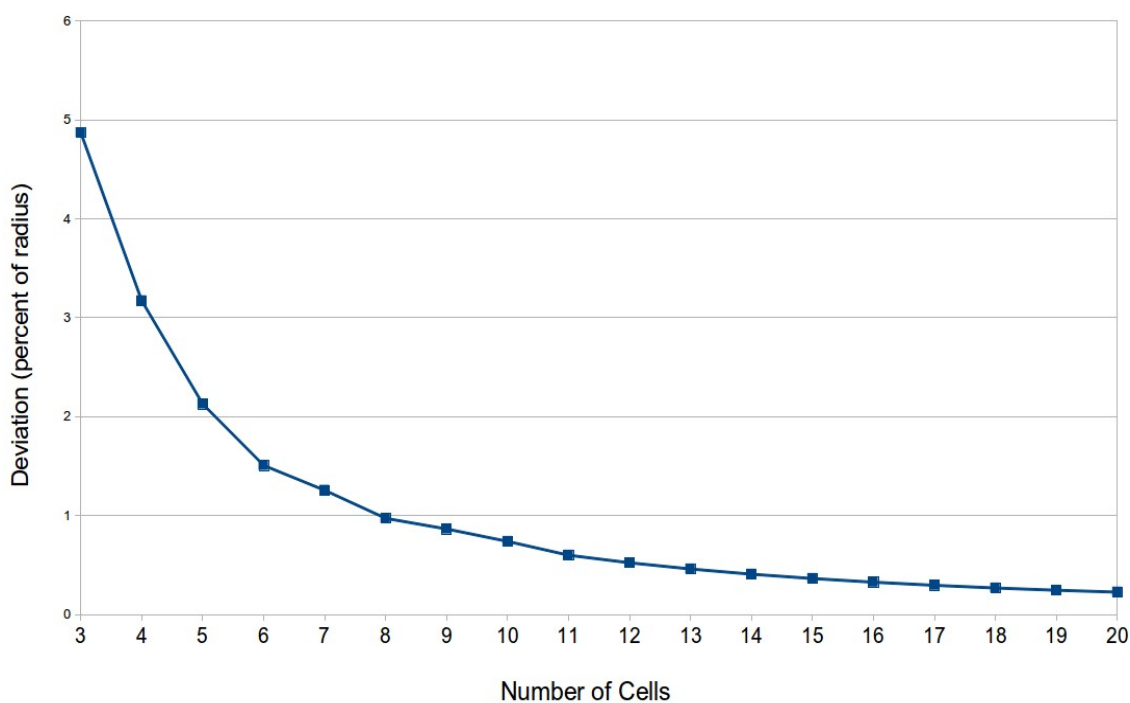

*Fig 10: Circularity increases with cell count.*

## 7.1  Orbiting strings can be redirected

Adding one cell near the 3-cell orbiting string represents the simplest scenario in which a string that would typically produce a hexagonal orbit in 60 steps, instead 'orbits' around the cell at (-1, 1, -2) in 132 steps (see Fig 11)[3]. By adding another cell at (-4, 4, -2)[4], we further modify the string's path, so that it now takes 204 steps to complete its circuit around the two 'guide' cells. We can guide this string in different directions, such as by adding another cell at (-4, 7, -5)[5].

---

1  http://busyboxes.org/n coordinates: (0,2,-1),(2,0,1),(2,2,0)
2  http://busyboxes.org/o coordinates: (0,2,-1),(2,0,1),(2,2,0),(2,-1,3),(2,-3,4),(2,-4,6),(2,-6,7),(2,-7,9)
3  http://busyboxes.org/p coordinates: (1,-1,-5),(2,1,-5),(3,2,-4),(-1,1,-2)
4  http://busyboxes.org/q coordinates: (1,-1,-5),(2,1,-5),(3,2,-4),(-1,1,-2),(-4,4,-2)
5  http://busyboxes.org/r coordinates: (1,-1,-5),(2,1,-5),(3,2,-4),(-1,1,-2),(-4,4,-2),(-4,7,-5)

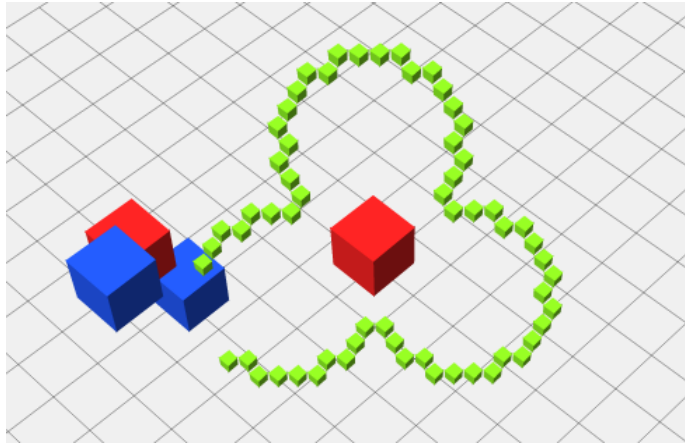
*Fig 11: string orbiting around a cell*

## 8   What goes in, must come out: implications of reversibility

We have found configurations of cells that exhibit extremely long periods of seemingly random behavior. The longest periods result in a cycle, where the configuration repeats itself after some number of time steps. However, in some cases the configuration 'breaks down', emitting one or more moving strings, and transitioning to a configuration of fewer cells that is either static or repeats, typically with a shorter cycle time. It is not obvious whether there exist analytic solutions to the question of whether a given configuration of cells will eventually cycle, or break down in some way.

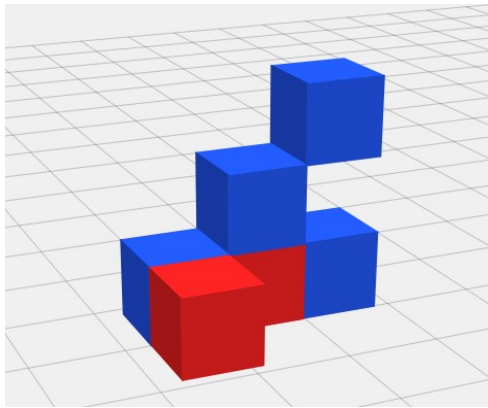
*Fig 12: 6-cell configuration cycles after 1200 steps*

Fig 12 shows a six-cell configuration that cycles after 1200 iterations[1]. One 13-cell configuration goes through over 6 million iterations before repeating[2]. A 14-cell cycling configuration iterates over 83 million steps[3], and there is a 15-cell configuration that iterates for at least 140 million steps, after which its fate is uncertain[4].

---

1   http://busyboxes.org/s coordinates: (-1,-1,-1),(-1,-1,0),(-1,-1,1),(-1,0,0),(-1,1,1),(0,-1,-1)
2   http://busyboxes.org/t coordinates: (-1,-2,0),(-1,1,1),(-2,-2,-1),(-2,-2,-2),(-2,0,1),(-3,-2,0),(0,1,-2),(1,-1,-1),(1,-1,1),(1,-2,-2),(1,0,-1), (1,1,-1), (1,1,0)
3   http://busyboxes.org/u coordinates: (-1,-1,-1),(-1,1,1),(-2,-1,-1),(-2,-1,-2),(-2,-2,-1),(-2,-3,0),(-2,1,0),(0,-2,1),(1,-1,-3),(1,1,-1),(1,1,1), (1,2,-1), (2,1,-1),(2,1,-2)
4   http://busyboxes.org/v coordinates: (-2,-1,-1),(-2,-1,-2),(-2,-2,-1),(-2,-2,1),(-2,0,-1),(-2,0,1),(-2,1,0),(0,-2,-1),(0,-2,1),(0,1,-1),(0,1,-2), (1,-2,-2),(1,0,-2),(1,1,-1),(1,1,0)

One 10-cell configuration[1] cycles after 229,110 iterations, whereas a similar 10-cell configuration evolves for almost 9,000 steps before emitting a 2-cell string, leaving an 8-cell configuration with a 36-step cycle[2]. It is interesting to note that due to the reversible nature of the Busy Boxes CA, we can make certain observations about past and future behavior in a case such as this. If we run the same configuration in reverse, we see that it too emits a string and 'decays' into a different 8-cell configuration, with a cycle time of 432[3]. We can summarize this sort of behavior with the rule, *what goes in, must come out:* In a reversible, discrete system of this sort (with finite bounds), it is impossible, from within the system, for a series of collisions to create a pattern which, if undisturbed, will continue to evolve without ever emitting a string or other glider, or disintegrating in some other way. In a time-symmetrical sense, we can also say that, given a pattern that does not last forever (ie it spontaneously 'decays'), we can know for certain that it must have undergone a prior emission or interaction with other elements at some time in the past.

The reason we can make these claims is as follows: as the system evolves in a finite implementation of a reversible CA, the state of every encountered configuration of cells must be part of the same single cycle. If we assume a finite configuration that spontaneously emits a particle, we can prove that it also had a prior interaction. If it didn't, then in the reverse direction it must enter a repetitive cycle since it is finite and deterministic. If it cycles in the reverse direction, that means that in the forward direction it must also cycle. Thus, we have a contradiction - Q. E. D.

There can also exist patterns that are neither created nor destroyed, but are simply 'there' throughout the time cycle[4]. Such patterns can be referred to as *singularities;* they exist by caveat, ie they are stipulated as part of the initial (and final!) conditions. So far, the only such patterns we are aware of are static; it is an open question whether a singularity pattern exists which exhibits any complex behavior, such as a repeating cycle or a cycle with a translation (ie, a glider).

## 9   Particle Creation and Decay

Given a random set of initial conditions, we have observed interesting behavior, including the formation and 'decay' of configurations such as the 3-cell orbiter shown in Fig. 8. We have run simulations on a 32 x 32 x 32 cubic grid, with 'wrap' mode enabled, so that each face of the cube is adjacent to the opposite face, forming a 4-dimensional toroidal topology (the important point being that there are no anomalous boundaries; a glider in an empty space of this kind will travel indefinitely).

We initialize our space with 200 *up* cells, 100 *even* and 100 *odd*, at locations chosen at random. We allow the system to evolve for 50,000 generations, in order for it to reach an equilibrium state of gliders and stationary cells. At that point, we start to keep track of the creation and destruction of the 3-cell orbiting pattern, in any orientation or reflection. If the pattern persists, it will orbit with a 60-generation cycle. Every 60 generations, we check to see if the pattern is still where we first located it. If it is, we count that as 'survival'; should the pattern not appear, we count the previous observation as the end of its lifetime.

---

1 http://busyboxes.org/w coordinates: (-2,1,-2),(-2,1,0),(0,-1,3),(0,0,1),(0,0,3),(0,1,-1),(0,1,0),(0,1,1),(0,4,1),(2,1,0)
2 http://busyboxes.org/x coordinates: (-1,1,-1),(0,-2,0),(0,0,3),(0,0,4),(0,1,-1),(0,1,0),(0,1,1),(0,4,1),(1,0,0),(1,1,1)
3 http://busyboxes.org/y coordinates: (-1,1,-1),(0,-2,0),(0,0,3),(0,0,4),(0,1,-1),(0,1,0),(0,1,1),(0,4,1),(1,0,0),(1,1,1)
4 http://busyboxes.org/z coordinates: (-1,0,0),(-1,0,1),(-1,1,0),(-1,1,1),(0,0,0),(0,0,1),(0,1,0),(0,1,1)

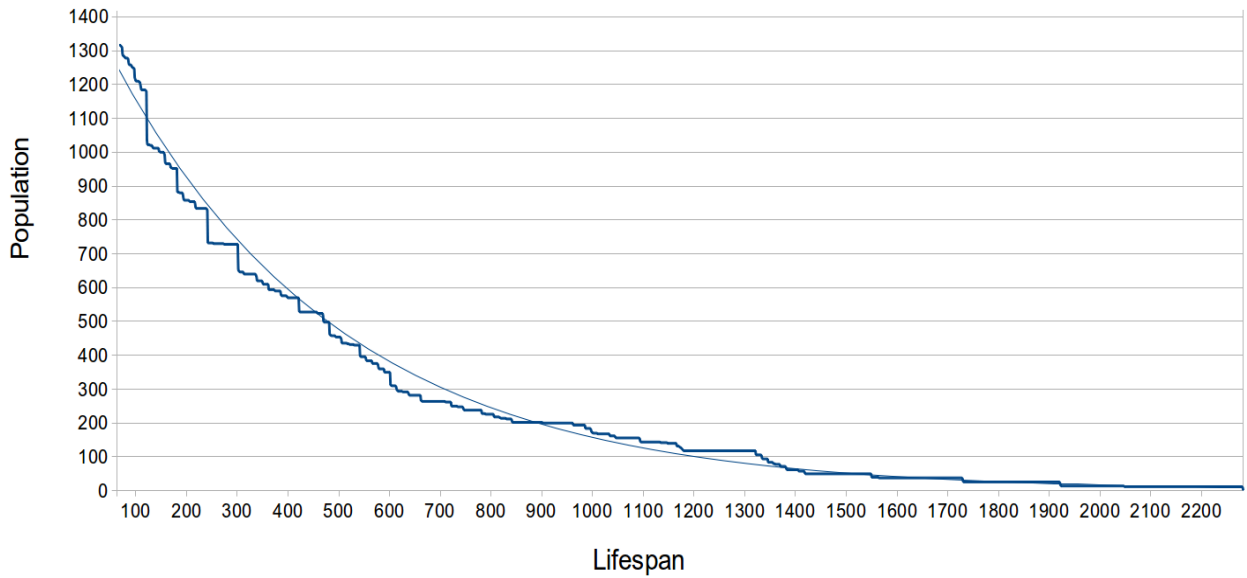

*Fig 13: population plot of 3-cell configuration[9] with exponential regression trend line*

In Fig. 13, we have plotted the lifetime of each such pattern that survived more than 60 generations, detected over an 800,000 generation run. The X axis represents the lifetime, and the Y axis represents the number of instances that lived at least that long.[1] The data shows a typical half-life decay pattern, as evidenced by the exponential regression trend line, shown in light blue.

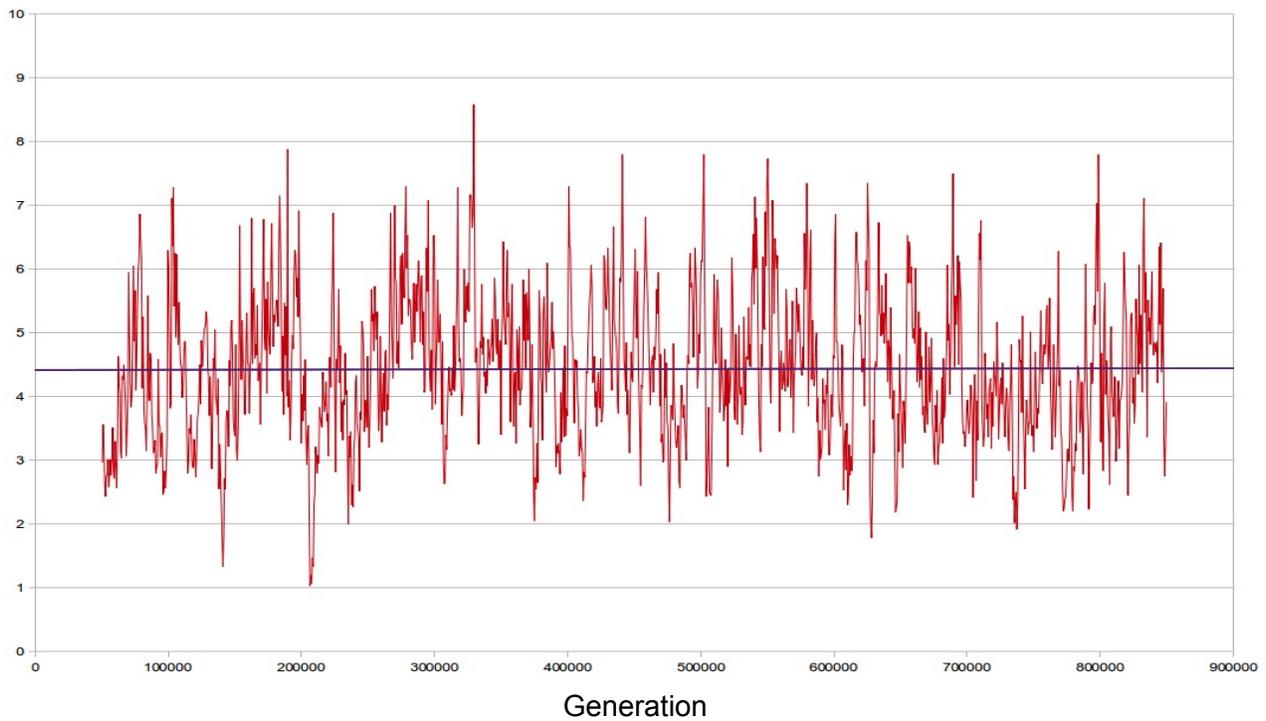

*Fig 14: average number of active cells during decay trial*

---

1   It should be noted that due to limitations of the software, a cycling orbiter is typically detected as 12 instances of the pattern as it moves through its cycle. Each instance is tracked separately.

Fig. 14 shows a plot of the number of 'active' cells -- cells that are swapped with a cell of a different state (ie cells that move), over the time period of the 800,000 generation run mentioned previously.  The dark purple line is a linear regression of the data; the fact that this line is nearly horizontal shows that the number of active cells is in equilibrium over the time period, and therefore is not a factor in the decay time measured.

## 10   Conclusion

The Busy Boxes variant of the SALT family of reversible cellular automata has shown itself to possess an intriguing number of features that are worthy of further study and analysis. Physicist Gerard T'Hooft has written about the possibility that reversible CA's such as this one may help elucidate some aspects of physics in the quantum realm [5]. We believe that the study of emergent, qualitative, and statistically predictable behavior of these kinds of discrete, deterministic systems can indeed shed light on questions in physics, and may also help to clarify the deeper principles of entropy and the origin of complex, self-propagating systems such as those found in our biosphere [6] [7] [8].